\makeatletter
\newcommand{\dontusepackage}[2][]{%
  \@namedef{ver@#2.sty}{9999/12/31}%
  \@namedef{opt@#2.sty}{#1}}
\makeatother
\dontusepackage{subfigure}

\documentclass[]{article}

\usepackage{lmodern}
\usepackage{amssymb,amsmath}
\usepackage{ifxetex,ifluatex}
\usepackage[usenames,dvipsnames]{color}
\usepackage{fixltx2e} 
\ifnum 0\ifxetex 1\fi\ifluatex 1\fi=0 
  \usepackage[T1]{fontenc}
  \usepackage[utf8]{inputenc}
\else 
  \ifxetex
    \usepackage{mathspec}
    \usepackage{xltxtra,xunicode}
  \else
    \usepackage{fontspec}
  \fi
  \defaultfontfeatures{Mapping=tex-text,Scale=MatchLowercase}
  
\fi
\IfFileExists{upquote.sty}{\usepackage{upquote}}{}
\IfFileExists{microtype.sty}{%
\usepackage{microtype}
\UseMicrotypeSet[protrusion]{basicmath} 
}{}
\usepackage[margin=1.0in,bottom=1.5in]{geometry}
\usepackage[square,numbers,sort&compress]{natbib}
\usepackage{listings}
\usepackage{graphicx}
\makeatletter
\def\maxwidth{\ifdim\Gin@nat@width>\linewidth\linewidth\else\Gin@nat@width\fi}
\def\maxheight{\ifdim\Gin@nat@height>\textheight\textheight\else\Gin@nat@height\fi}
\makeatother
\setkeys{Gin}{width=\maxwidth,height=\maxheight,keepaspectratio}
\usepackage{caption}
\usepackage{float}
\usepackage{subfig}
\ifxetex
  \usepackage[setpagesize=false, 
              unicode=false, 
              xetex]{hyperref}
\else
  \usepackage[unicode=true]{hyperref}
\fi
\hypersetup{breaklinks=true,
            bookmarks=true,
            pdfauthor={},
            pdftitle={Learning by example\{:\} fast reliability-aware seismic imaging with normalizing flows},
            colorlinks=true,
            citecolor=black,
            urlcolor=blue,
            linkcolor=black,
            pdfborder={0 0 0}}
\usepackage[preprint]{neurips_2019}
\usepackage{url}
\usepackage{booktabs}
\usepackage{amsfonts}
\usepackage{nicefrac}
\usepackage{microtype}

\usepackage{mathtools}
\newcommand{\bx}{\mathbf{x}}
\newcommand{\rx}{\mathrm{x}}
\newcommand{\ry}{\mathrm{y}}
\newcommand{\by}{\mathbf{y}}
\newcommand{\bz}{\mathbf{z}}
\newcommand{\rz}{\mathrm{z}}
\newcommand{\bw}{\mathbf{w}}
\newcommand{\rw}{\mathrm{w}}

\title{Learning by example$\boldsymbol{:}$ fast reliability-aware seismic
imaging with normalizing flows}
\author{Ali Siahkoohi and Felix J. Herrmann\\School of Computational Science and
Engineering,\\Georgia Institute of
Technology\\\texttt{\{alisk,\phantom{\ }felix.herrmann\}@gatech.edu}}
\date{}

\begin{document}
\maketitle
\begin{abstract}
Uncertainty quantification provides quantitative measures on the
reliability of candidate solutions of ill-posed inverse problems. Due to
their sequential nature, Monte Carlo sampling methods require large
numbers of sampling steps for accurate Bayesian inference and are often
computationally infeasible for large-scale inverse problems, such as
seismic imaging. Our main contribution is a data-driven variational
inference approach where we train a normalizing flow (NF), a type of
invertible neural net, capable of cheaply sampling the posterior
distribution given previously unseen seismic data from neighboring
surveys. To arrive at this result, we train the NF on pairs of low- and
high-fidelity migrated images. In our numerical example, we obtain
high-fidelity images from the Parihaka dataset and low-fidelity images
are derived from these images through the process of demigration,
followed by adding noise and migration. During inference, given shot
records from a new neighboring seismic survey, we first compute the
reverse-time migration image. Next, by feeding this low-fidelity
migrated image to the NF we gain access to samples from the posterior
distribution virtually for free. We use these samples to compute a
high-fidelity image including a first assessment of the image's
reliability. To our knowledge, this is the first attempt to train a
conditional network on what we know from neighboring images to improve
the current image and assess its reliability.
\end{abstract}

\section{Introduction}\label{introduction}

Seismic imaging is challenged by noise and a non-trivial null-space due
to bandwidth and aperture limitations. These challenges lead to solution
non-uniqueness where different images may fit the data equally well.
Solution non-uniqueness can be characterized via a distribution over the
solution space that assigns probabilities to different estimates---i.e.,
the posterior distribution \citep{tarantola2005inverse}. Due to the
high-dimensionality of seismic images, posterior inference often
requires high-dimensional sampling, for instance via Markov chain Monte
Carlo \citep[MCMC,][]{robert2004monte} techniques. MCMC methods sample
the posterior distribution via a series of random walks in the
probability space where the posterior probability density function (PDF)
needs to be evaluated or approximated at each step---e.g., via
stochastic gradient Langevin dynamics \citep{welling2011bayesian}. These
sampling methods typically require many steps to traverse the
probability space
\citep{malinverno2004expanded, malinverno2006two, MartinMcMC2012, chevron2017, fang2018uqfip, stuart2019two, zhao2019gradient, kotsi2020},
which fundamentally limits their applicability to large-scale problems
due to costs associated with the forward operator
\citep{curtis2001prior, herrmann2019NIPSliwcuc, siahkoohi2020EAGEdlb, siahkoohi2020SEGhorizonUQ}.
Alternatively, variational inference methods
\citep{jordan1999introduction} approximate the posterior distribution
with a parametric and easy-to-sample distribution. By virtue of this
approximation, sampling is turned into an optimization problem where the
parameters are tuned to minimize the ``distance'' between the parametric
and posterior distributions. After optimization, this easy-to-sample
distribution is used to conduct Bayesian inference. As discussed by
\citet{blei2017variational}, variational-inference based methods are
known to have computational advantages over MCMC in high-dimensional
problems.

In this paper, we take a variational inference approach to seismic
imaging and use a normalizing flow \citep[NF,][]{rezende2015variational}
as a parametric function to approximate the posterior distribution. NFs
are a special type of invertible neural nets
\citep{dinh2016density, lensink2019fully, kruse2019hint} capable of
approximating complicated functions---e.g., non-Gaussian PDFs. After
training, NFs cheaply sample the distribution of interest and estimate
its PDF. The latter property sets NFs apart from generative adversarial
nets \citep[GANs,][]{Goodfellow2014} since GANs only provide samples
from the distribution of interest and are incapable of explicitly
representing densities by design \citep{Goodfellow2014}. Despite this
disadvantage, GANs have been used successfully to carry out stochastic
waveform inversions with the Metropolis-adjusted Langevin algorithm
\citep{Mosser2020}. Contrary to GANs, NFs can be simply trained via the
maximum-likelihood estimation method \citep{rezende2015variational},
which is asymptotically consistent and efficient
\citep{bishop2006pattern}. On the other hand, training GANs typically
involves a min-max optimization that might lead to unstable training
\citep{wgan}. Finally, the invertibility of NFs allows for
memory-efficient training since the intermediate state variables do not
need to be stored \citep{Putzky2019, peters2020fully, invnet}.

We use a data-driven approach for training the NF to take maximum
advantage of available data in the form of low- and high-fidelity
migrated images. One way of creating training data is to pair
high-fidelity least-squares migration images with low-fidelity
reverse-time migration images from nearby surveys. In this work, we
mimic this choice by pairing high-fidelity images from the
\href{https://wiki.seg.org/wiki/Parihaka-3D}{Parihaka} dataset
\citep{WesternGeco2012} with low-fidelity images obtained via the
process of demigration, followed by adding noise and migration. This
approach circumvents the challenges of explicitly choosing a prior
distribution that captures the true heterogeneity exhibited by the
Earth's subsurface. Moreover, the NF trained via this method is not
specific to one seismic survey, but thanks to the NF's ability to
generalize can be used to characterize the posterior distribution
associated with a previously unseen seismic survey, which is in some
sense close---e.g., data from a neighboring survey area. Finally, this
approach, unlike MCMC methods, does not involve repeated evaluations of
the forward operator, however, by providing the reverse-time migrated
image to the NF we implicitly capture the likelihood and prior in our
learned posterior distribution. To our knowledge, this is the first
normalizing-flow based scalable attempt to cheaply sample from the
seismic imaging posterior distribution given previously unseen seismic
data.

In the context of variational inference for inverse problems with
expensive forward operators, Stein variational gradient descent
\citep{NIPS2017_17ed8abe} is used to iteratively push samples from the
prior to the posterior distribution to tackle full-waveform inversion
\citep{zhang2020seismic} and seismic tomography \citep{Zhang2020}.
\citet{rizzuti2020SEGuqavp} and \citet{zhao2020bayesian} both
independently proposed a physics-based variational inference approach
that approximates the posterior distribution with a distribution
represented by a NF. While this approach does not require training data,
it does need choosing a prior and repeated computationally expensive
evaluations of the forward operator and the adjoint of its Jacobian
(migration). However, there are indications that this formulation can be
faster and easier to scale to large-scale problems compared to MCMC
methods \citep{zhao2020bayesian}. Motivated by \citet{kruse2019hint},
\citet{siahkoohi2020ABIpto} take this approach a step further by
proposing a two-stage multifidelity approach where during the first
stage a conditional NF is trained on available pairs of low- and
high-fidelity images. This pretrained NF is subsequently fine tuned
during an optional second physics-based variational inference stage,
which is tailored to the physics of the specific imaging problem at
hand. In this work, we focus on the first stage because it is
computational cheap and provides a rudimentary assessment of the image's
reliability. Moreover, since the first stage entails training over low-
and high-fidelity image pairs of related imaging problems, it can serve
as a warm start for the optional second more accurate physics-based
inference stage \citep{siahkoohi2020TRfuqf, siahkoohi2020ABIpto},
greatly reducing its costs. In related data-driven GAN-based approaches,
\citet{adler2018deep} and \citet{kovachki2020conditional} also train
neural nets capable of sampling from the posterior distribution given
new data. Instead of adhering to GANs, we rely on NFs, which are easier
to train and have a much more limited memory footprint during the
training phase, making them applicable to larger-scale problems.

In the following section, we first briefly introduce variational
inference and mathematically state the foundations of NFs. Next, we
describe our proposed normalizing-flow based formulation for seismic
imaging and Bayesian inference. We conclude by showcasing our approach
on a ``quasi'' real example that derives from 2D subsets of the Parihaka
dataset.

\section{Variational inference}\label{variational-inference}

At its core, the method we are proposing characterizes multivariate
distributions given access to samples from these distributions only.
Variational inference provides the statistical framework for this and
involves the problem of approximating a distribution of interest $p$
with a parametric distribution $q$, represented in our case by a NF.
This approximation is typically achieved by minimizing the
Kullback-Leibler (KL) divergence between $q$ and $p$. The KL divergence
offers a distance metric for distributions and can be explained as the
cross-entropy of distribution $p$ relative to distribution $q$ minus the
entropy of distribution $q$. During variational inference, the KL
divergence is minimized with respect to distribution $q$. By design, $q$
is typically chosen to be easy---i.e., computationally cheap, to sample
from. As a result, after optimization, we can cheaply sample from the
distribution $p$ by sampling $q$ instead. Next, we introduce NFs as a
way to parameterize distribution $q$.

\subsection{Normalizing flows}\label{normalizing-flows}

A NF $T_{\rw} : {X} \rightarrow {Z}$, parameterized by the NF weights
$\bw$, is an invertible neural net---i.e., it has a closed-from and
exact inverse (up to numerical precision), with the goal of mapping
inputs $\bx \in {X}$ sampled from the distribution of interest,
$\bx \sim p(\bx)$, into samples from the Gaussian distribution
$\bz \sim p_{\rz}(\bz)$, where $\bz \in {Z}$ and
$p_{\rz} (\bz) := \mathrm{N} (\bz \mid \mathbf{0}, \mathbf{I})$. In
other words, given samples $\{\bx^{(i)}\}_{i=1}^n \sim p(\bx)$, the
parameters $\bw$ are optimized so that the KL divergence between
$q_{\rw}$ and $p_z$ is minimized, where $q_{\rw}$ is the distribution of
the NF's output when $\bx \sim p(\bx)$ are fed as input. This mapping is
useful because after training we can sample from $p(\bx)$ by simply
inputting Gaussian samples $\bz \sim p_{\rz}(\bz)$ into the inverted
network, $T_{\rw}^{-1}$, provided that $q_{\rw}$ approximates $p_z$ well
\citep{rezende2015variational}. Given samples $\{\bx^{(i)}\}_{i=1}^n$
from the unknown distribution, minimizing the KL divergence between
$q_{\rw}$ and $p_z$ leads to the following optimization problem for
training the NF $T_{\rw}$ \citep{rezende2015variational}:
\begin{equation}
\mathop{\rm min}_{\bw}\, \frac{1}{n}
    \sum_{i=1}^n \Big[ \frac{1}{2}\big\| T_{\rw} (\bx^{(i)}) \big\|^2_2
    - \log \Big | \det \nabla_{\rx} T_{\rw} (\bx^{(i)}) \Big | \Big].
\label{NF-training}
\end{equation}
 In the above objective, the $\ell_2$-norm on $T_w$ imposes a Gaussian
distribution on the network output, which is what a NF is designed to
do---i.e.~to Gaussianize the input, $\bx^{(i)}$. The second term
involving $\det \nabla_{\rx} T_{\rw} (\bx)$ regularizes the entropy of
$q_{\rw}$, which avoids $T_w (\bx)$ from producing a trivial
output---i.e., $T_w (\bx) := \mathbf{0}$. The extra costs of computing
$\det \nabla_{\rx} T_{\rw} (\bx)$ and its gradient are negligible due to
the special design of invertible nets \citep{dinh2016density}. We solve
the optimization problem in Equation~\ref{NF-training} via stochastic
optimization where at each iteration the objective is approximated via
randomly selected (without replacement) training samples $\bx^{(i)}$. In
addition to sampling from $p(\bx)$ after training, we can also use the
change-of-variables formula \citep{villani2009optimal} to cheaply
estimate $p(\bx)$ for a given $\bx$ by
$p (\bx) \approx p_{\rz} (T_{\rw} (\bx))\, \Big | \det \nabla_{\rx} T_{\rw} (\bx) \Big |$.
In cases where we only have access to samples from the prior
distribution, this relation provides an expression for the prior PDF
that can be used in inverse problems involving estimating $\bx$.

\subsection{Posterior inference with conditional
NFs}\label{posterior-inference-with-conditional-nfs}

Our ultimate goal is to capture the posterior distribution associated
with inverse problems involving computationally expensive and possibly
nonlinear forward operators, $F: X \rightarrow Y$, with the forward
model, $\by = F (\bx) + \mathbf{n}$. Here, $\bx \in X$ represents the
unknown model, $\by \in Y$ the observed data, and $\mathbf{n}$ possibly
non-Gaussian noise. Following \citet{kruse2019hint}, NFs can be adapted
to sample from the posterior distribution, $p(\bx \mid \by)$ given
previously unseen data $\by$. NFs capable of generating samples from
conditional distributions can be defined by imposing a block-triangular
structure on $T_{\rw} : Y \times X \rightarrow Z \times Z$, which
according to \citet{kruse2019hint} leads to the following NF structure:
\begin{equation}
\begin{split}
T_{\rw} (\by, \bx)
    =\begin{bmatrix} T_{\rw_1} (\by) \\[4pt] T_{\rw_2}
    (\by, \bx) \end{bmatrix}, \
    \bw = \begin{bmatrix} \bw_1 \\ \bw_2 \end{bmatrix}.
\end{split}
\label{block-triangular}
\end{equation}
 In words, Equation~\ref{block-triangular} describes $T_{\rw}$ as a NF
that maps the joint random variable $(\by, \bx)$ into two Gaussian
random vectors, where the first output is only a function of $\by$. As
in Equation~\ref{NF-training}, this conditional NF can be trained given
training pairs, $\{\by^{(i)}, \bx^{(i)}\}_{i=1}^n$, by minimizing the
following objective:
\begin{equation}
\mathop{\rm min}_{\bw}\, \frac{1}{n}
    \sum_{i=1}^n \Big[ \frac{1}{2}\big\| T_{\rw} (\by^{(i)}, \bx^{(i)}) \big\|^2_2
    - \log \Big | \det \nabla_{(\ry, \rx)} T_{\rw} (\by^{(i)}, \bx^{(i)}) \Big | \Big].
\label{NF-training-cond}
\end{equation}
 This block-diagonal construction lets us sample from the posterior
distribution via \citep{marzouk2016sampling, kruse2019hint}
\begin{equation}
T_{\rw_2}^{-1} \big( T_{\rw_1} (\by), \bz \big) \sim p(\bx \mid \by), \quad \bz \sim p_{\rz}(\bz).
\label{cond-sampling}
\end{equation}
 According to this expression, samples from the posterior are drawn by
first by feeding the data, $\by$, to $T_{\rw_1}$. Next, posterior
samples are obtained by feeding $T_{\rw_1} (\by)$ and Gaussian samples
$\bz \sim p_{\rz}(\bz)$ into the inverse network, $T_{\rw_2}^{-1}$.
These Gaussian samples can be considered as realizations of the
posterior in the latent space. Compared to methods such as MCMC, this
sampling procedure does not involve the forward operator and is
therefore extremely fast and computationally cheap. Given samples
generated with Equation~\ref{cond-sampling}, we approximate the
conditional mean and point-wise standard deviation point-estimators via
\begin{equation}
\boldsymbol{\mu} = \mathbb{E} \big [ \bx \mid \by \big ], \quad \boldsymbol{\sigma}^2 = \mathbb{E} \big[ \left (\bx
    - \boldsymbol{\mu} \right ) ^{\circ 2} \mid \by \big ],
\label{point-est}
\end{equation}
 where $^{\circ 2}$ represents elementwise power of two and the
expectations can be approximated with sample means.

\subsection{Imaging with conditional
NFs}\label{imaging-with-conditional-nfs}

Seismic imaging is the problem of estimating the short-wavelength
squared-slowness perturbation model of the Earth's subsurface given
noisy observed seismic data and a kinematically accurate background
velocity model. We cast seismic imaging into the deep-learning based
framework described in the previous section to limit the computational
burden of high-dimensional posterior inference. To achieve this, we
choose $\bx$ to represent the high-fidelity migrated image and $\by$ the
corresponding low-fidelity reverse-time migrated image obtained by the
process of demigration, followed by adding noise and migration. With
this choice, the prior information in this technique comes from
different realizations of high-fidelity migrated images provided in
training pairs $(\by^{(i)}, \bx^{(i)}), \ i=1,\ldots,n$, obtained from
neighboring surveys. We train the NF by solving the optimization problem
in Equation~\ref{NF-training-cond} while using the block-triangular
structure (Equation~\ref{block-triangular}). This yields a conditional
NF that provides cheap, fast, and forward-operator free samples from the
imaging posterior distribution via Equation~\ref{cond-sampling}. We use
these samples to obtain a high-fidelity image including a first
assessment of its uncertainty.

\section{Numerical experiments}\label{numerical-experiments}

We propose a ``quasi'' real example in which we generate synthetic data
by applying the linearized Born scattering operator to $4750$ 2D
sections with size $3075\, \mathrm{m} \times 5120\, \mathrm{m}$
extracted from the real Kirchhoff migrated Parihaka dataset. We consider
a $12.5\, \mathrm{m}$ vertical and $20\, \mathrm{m}$ horizontal grid
spacing, and we augment a $125\, \mathrm{m}$ water column on top of
these models to limit the near source imaging artifacts. We parameterize
the linearized Born scattering operator via a made up smooth background
squared-slowness model. To ensure good coverage, we simulate $102$ shot
records with a source spacing of $50\, \mathrm{m}$. Each shot is
recorded for two seconds with $204$ fixed receivers sampled at
$25\, \mathrm{m}$ spread on top of the model. The source is a Ricker
wavelet with a central frequency of $30\, \mathrm{Hz}$. To mimic a more
realistic imaging scenario, we add band-limited noise to the shot
records, where the noise is obtained by filtering white noise with the
source wavelet. We create training pairs
$(\by^{(i)}, \bx^{(i)}), \ i=1,\ldots,4750$, by first simulating noisy
seismic data according to the above mentioned acquisition design for all
$\bx^{(i)}$. Next, we obtain $\by^{(i)}$ by applying
reverse-time-migration to the obtained data for each model $\bx^{(i)}$.
We train $T_w$ according to the objective function in
Equation~\ref{NF-training-cond} with the Adam \citep{kingma2015adam}
stochastic optimization method with batch size $16$ for $100$ passes
over the training dataset (epochs). We use an initial stepsize of
$10^{-4}$ and multiply it by $0.9$ every two epochs.

\begin{figure*}
\centering
\subfloat[\label{true-2}]{\includegraphics[width=0.500\hsize]{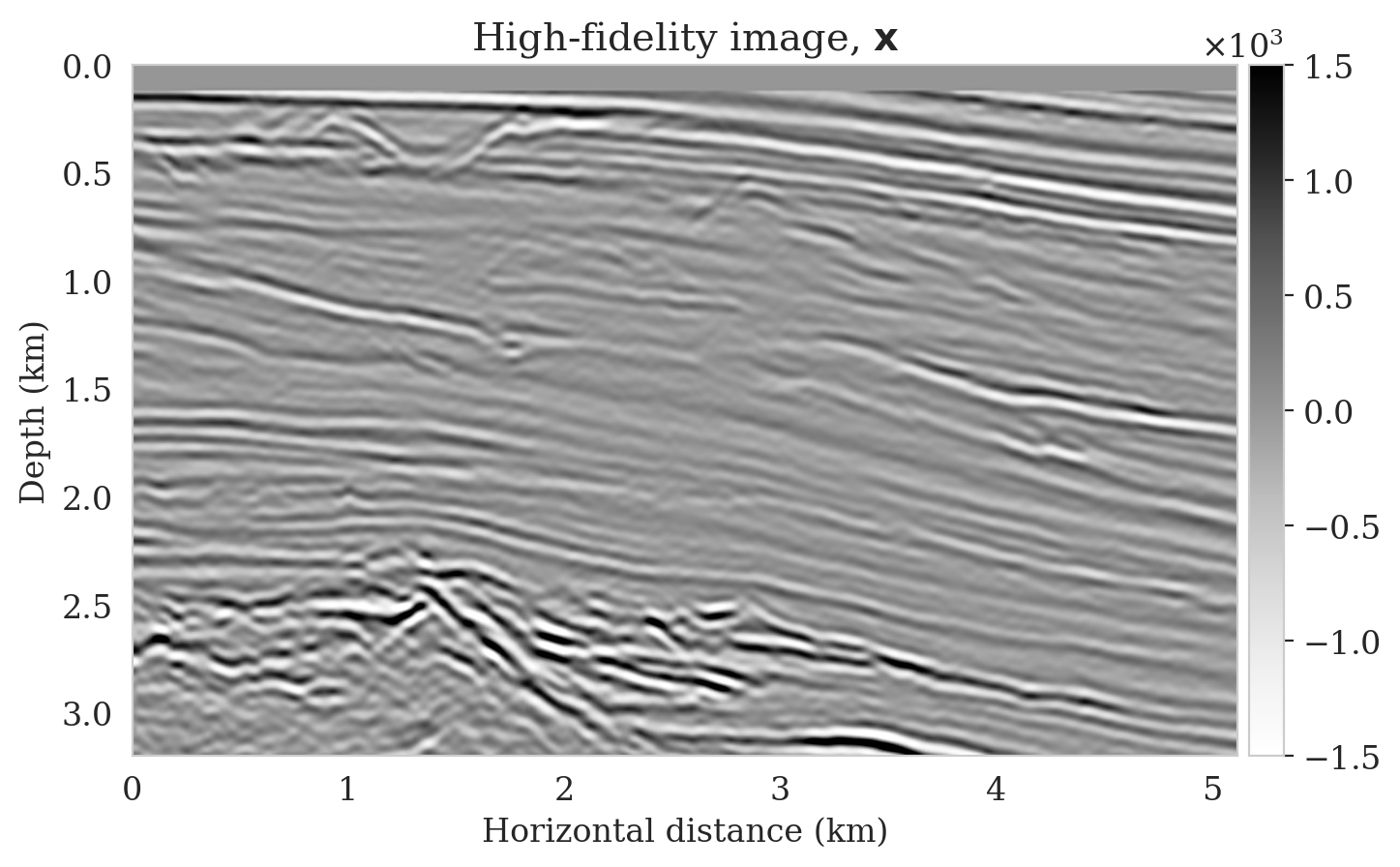}}
\subfloat[\label{rtm-2}]{\includegraphics[width=0.500\hsize]{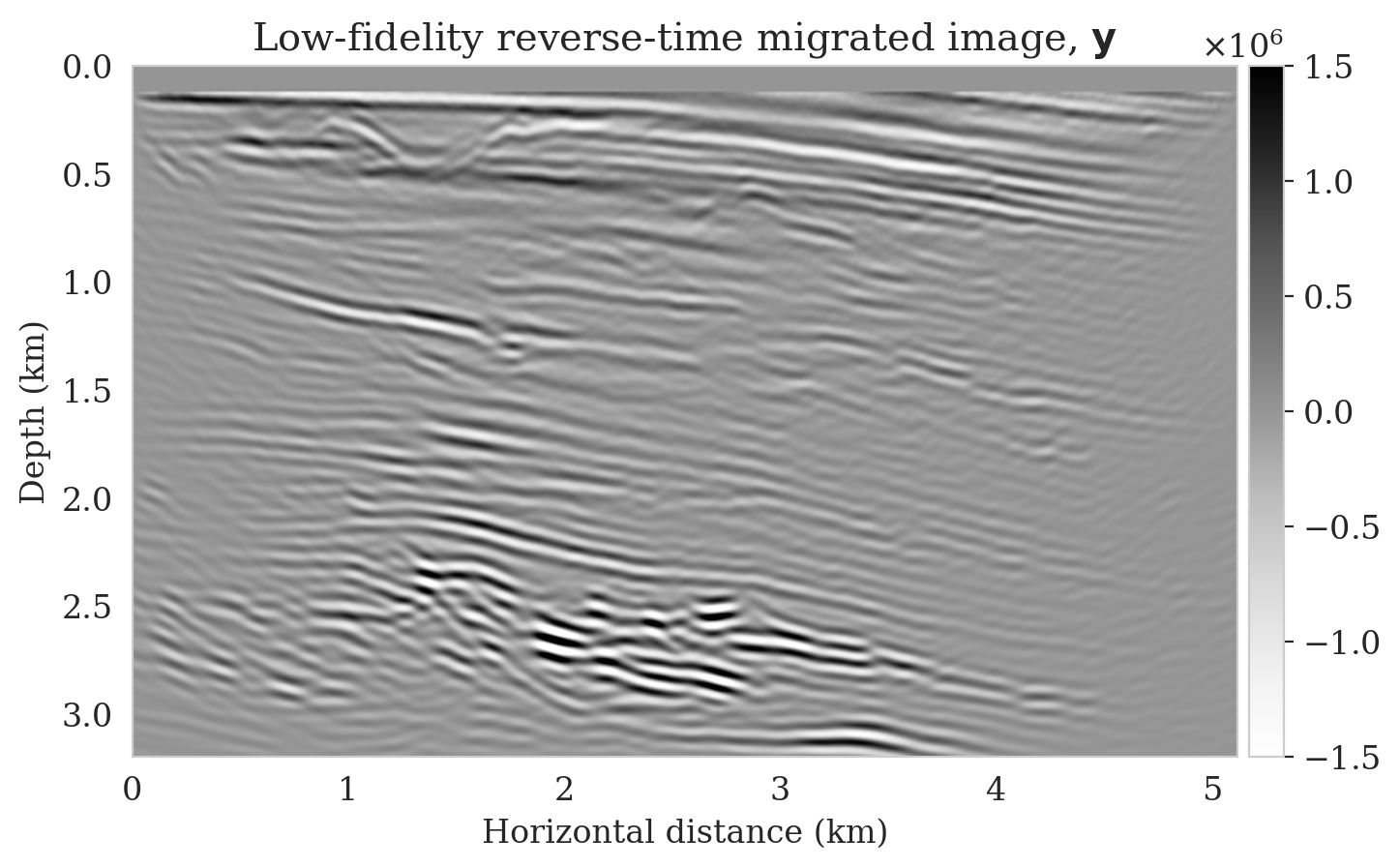}}
\\
\subfloat[\label{sample-2-1}]{\includegraphics[width=0.500\hsize]{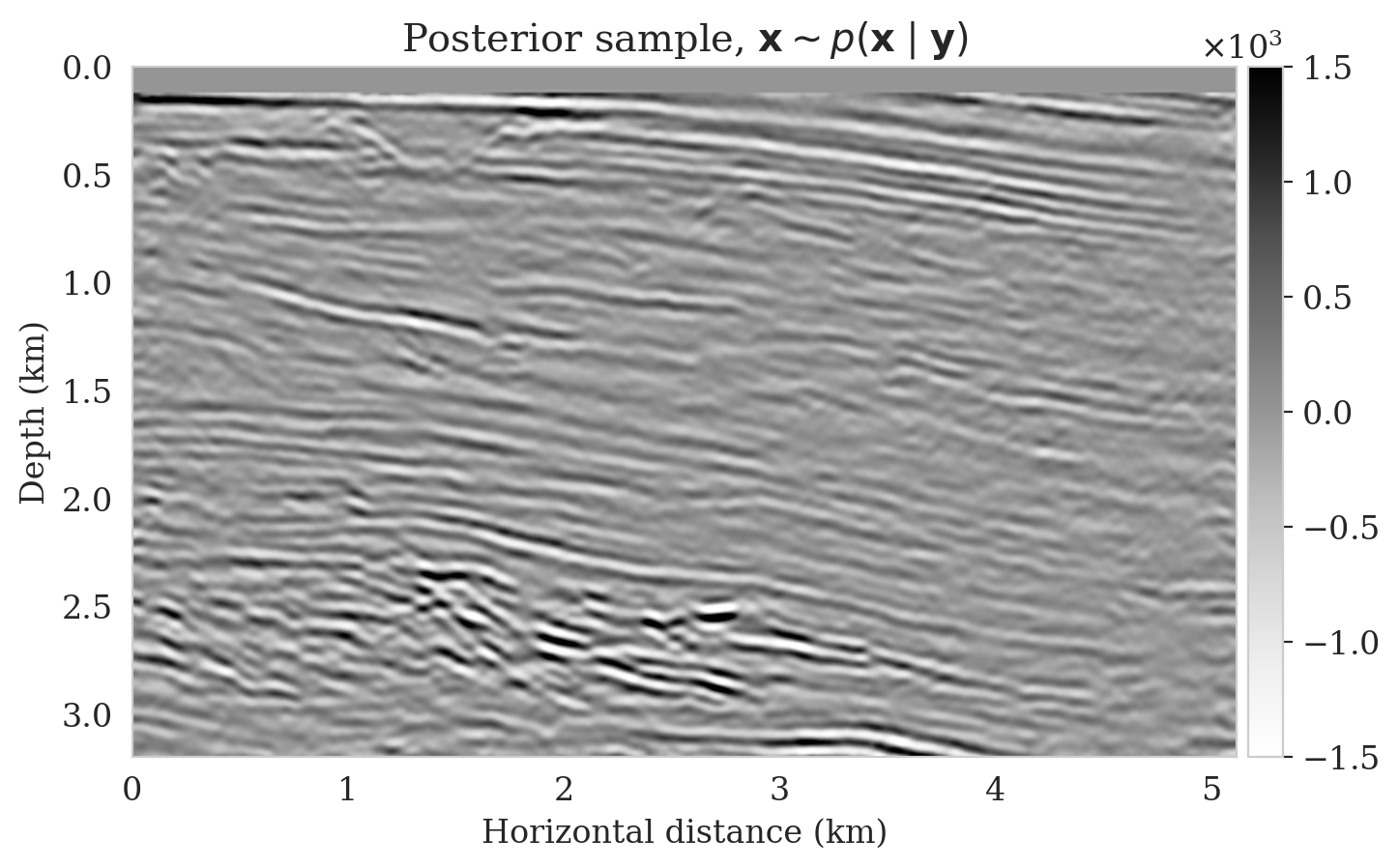}}
\subfloat[\label{sample-2-4}]{\includegraphics[width=0.500\hsize]{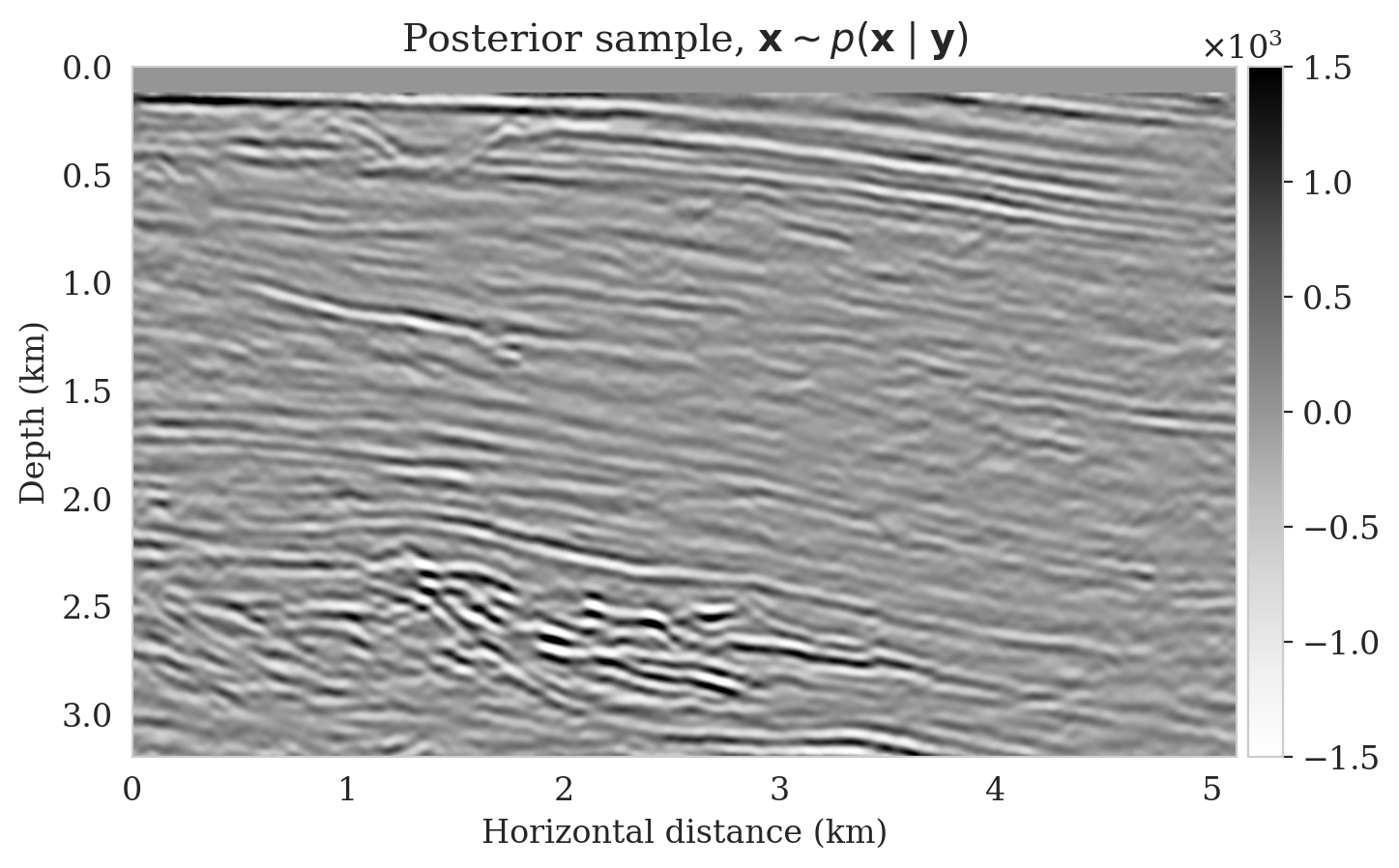}}
\\
\subfloat[\label{cm-2}]{\includegraphics[width=0.500\hsize]{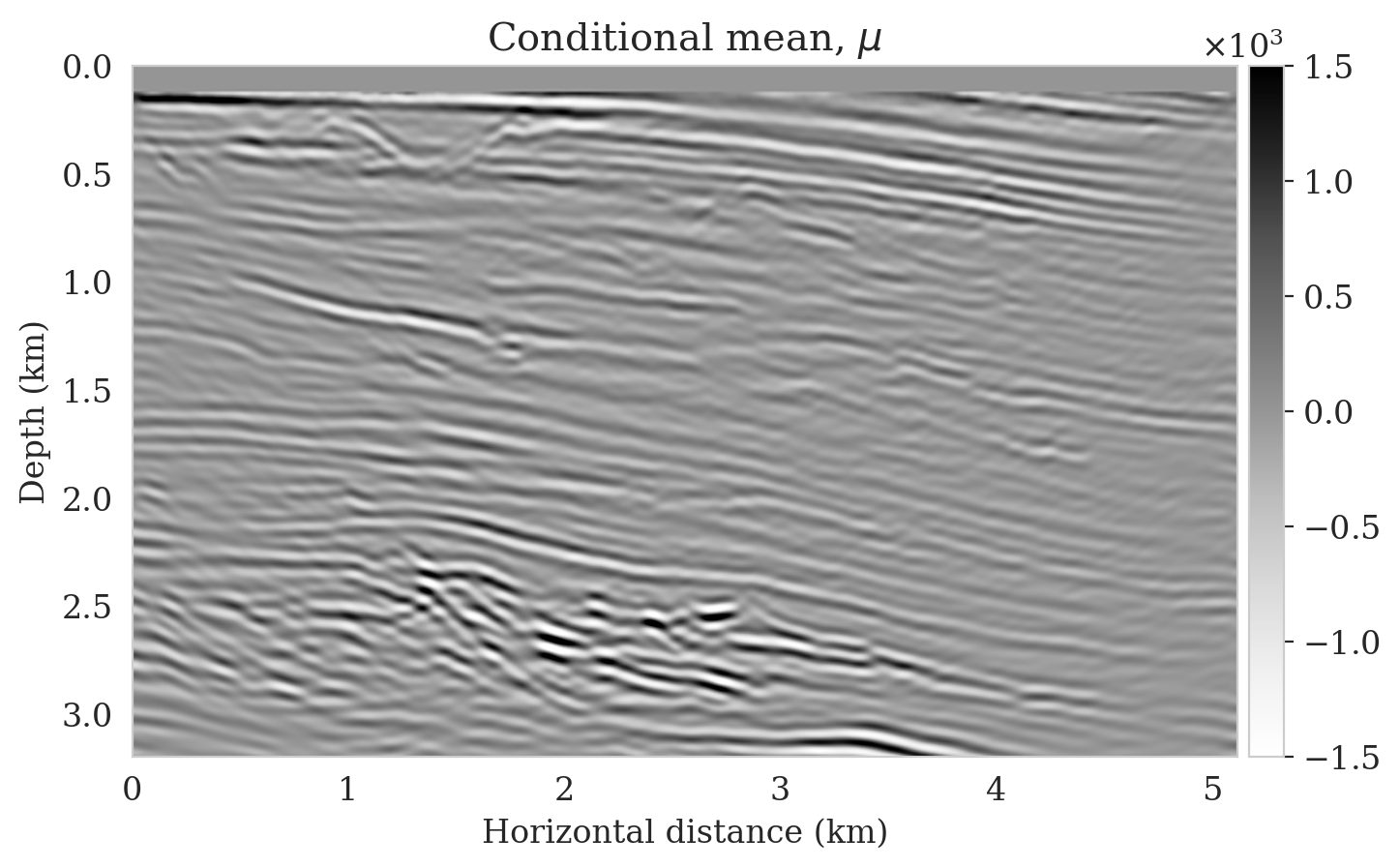}}
\subfloat[\label{std-2}]{\includegraphics[width=0.500\hsize]{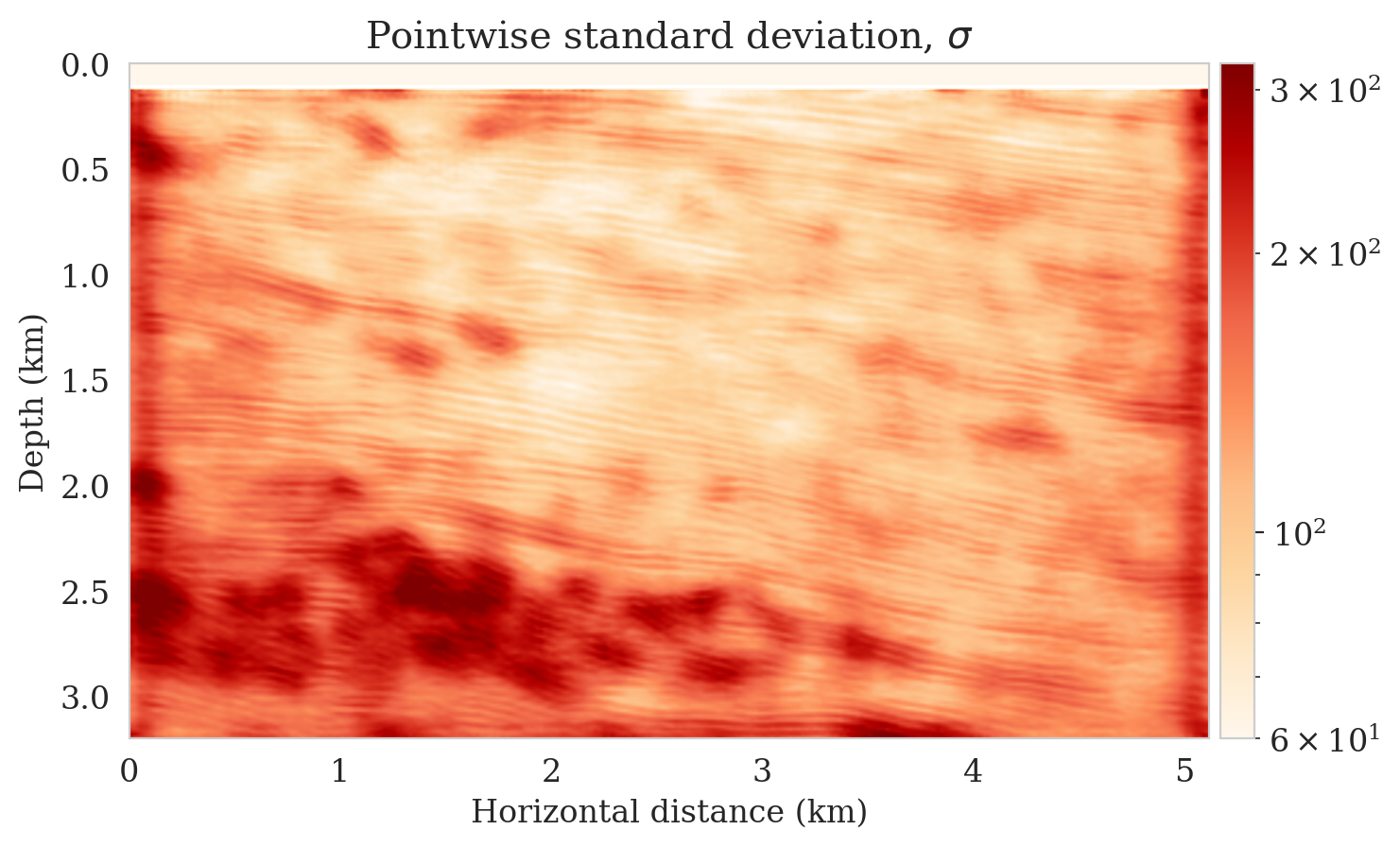}}
\caption{Reliability-aware imaging. (a) High-fidelity image. (b)
Reverse-time migrated image. (c) and (d) Samples from posterior
distribution. (e) Conditional mean and (f) pointwise standard deviation
via the proposed method according to
Equation~\ref{point-est}.}\label{example}
\end{figure*}

After training, we simulate noisy seismic data for a previously unseen
perturbation model extracted from the Parihaka dataset, shown in
Figure~\ref{true-2}. The corresponding low-fidelity reverse-time
migrated image is shown in Figure~\ref{rtm-2}. Given this image, we use
Equation~\ref{cond-sampling} to generate $1000$ random samples
$p(\bx \mid \by)$ at a negligible cost. Two of these samples are shown
in Figures~\ref{sample-2-1} and~\ref{sample-2-4}. With these $1000$
samples, we also estimate the conditional mean (Figure~\ref{cm-2}) and
pointwise standard deviation (Figure~\ref{std-2}). Because of its
reported \citep{anderson1979} statistical robustness, we use the
conditional mean as an estimate for the high-fidelity image. This
estimate includes pointwise standard deviations that serve as an
assessment of the uncertainty. From Figure~\ref{example}, we observe
that the overall amplitudes are well recovered by the conditional mean
estimate, which includes partially recovered reflectors in badly
illuminated areas close to the boundaries. Although the reconstructions
are not perfect, they significantly improve upon the reverse-time
migrations estimate. As expected, the pointwise standard deviation in
Figure~\ref{std-2} indicates that we have the most uncertainty in areas
of complex geology---e.g., near channels and tortuous reflectors, and in
areas with a relatively poor illumination (deep and close to
boundaries). The areas with large uncertainty align with
difficult-to-image parts of the model.

In this example, we use \href{https://www.devitoproject.org/}{Devito}
\citep{devito-compiler, devito-api} for the wave-equation based
simulations. For implementation of the network architectures, we rely on
\href{https://github.com/slimgroup/InvertibleNetworks.jl}{InvertibleNetworks.jl}
\citep{invnet}, a memory-efficient framework for training invertible
nets in the Julia programming language. Finally, code to reproduce our
results are made available on
\url{https://github.com/slimgroup/Software.SEG2021}.

\section{Conclusions and discussion}\label{conclusions-and-discussion}

Seismic imaging is challenging due to complex and computationally
expensive-to-evaluate forward operators and the highly heterogeneous
multiscale structure of the Earth. These challenges limit the
applicability of Markov Chain sampling methods due to the costs
associated with the forward operator. These difficulties are compounded
by difficulties in capturing the heterogeneity exhibited by the Earth's
subsurface in the form of a prior distribution. To handle this situation
and to assess uncertainty, we propose a data-driven, conditional
normalizing-flow based variational inference approach. The proposed
inference scheme takes advantage of having access to training
pairs---e.g., in form of high-fidelity seismic images and low-fidelity
remigrated images, to train a conditional normalizing flow capable of
sampling from the posterior distribution. Because neural nets are known
to generalize, the trained conditional normalizing flow can be used to
quickly generate samples from the posterior given unseen low-fidelity
images. In turn, these samples provide access to the high-fidelity
conditional mean estimate including a rudimentary assessment of its
uncertainty. In that sense, the presented imaging scheme learns by
example, maximally leveraging what is already known. Since we can not
completely rely on generalization of neural nets, the trained
conditional normalizing flow can be used as input to a second
physics-based stage of inference where the normalizing flow undergoes
transfer learning and then samples from the posterior specific to the
imaging problem at hand. By means of a numerical example, we demonstrate
that our proposed scheme indeed provides fast access to posterior
samples for problems that would otherwise be computationally unfeasible
to solve with the more traditional methods. Since the presented method
relies on generalization by conditional normalizing flows, we stop short
of claiming that we quantify uncertainty but rather that we supply a
measure of reliability of the estimated image.

\section{Acknowledgment}\label{acknowledgment}

This research was carried out with the support of Georgia Research
Alliance and partners of the ML4Seismic Center.

\bibliography{abstract}

\end{document}